\documentclass[12pt]{article}  
\pdfoutput=1

\newcommand{\xbar}{\mbox{$\vec{x}$}}
\newcommand{\xvec}{\mbox{$\vec{x}$}}
\newcommand{\cvec}{\mbox{$\vec{c}$}}

\newcommand{\nvec}{\mbox{$\vec{n}$}}
\newcommand{\muvec}{\mbox{$\vec{\mu}$}}

\newcommand{\xivec}{\mbox{$\vec{\xi}$}}
\newcommand{\chivec}{\mbox{$\vec{\chi}$}}
\newcommand{\Nvec}{\mbox{$\vec{N}$}}
\newcommand{\Mvec}{\mbox{$\vec{M}$}}

\newcommand{\Avec}{\mbox{$\vec{A}$}}
\newcommand{\Xvec}{\mbox{$\vec{X}$}}
\newcommand{\Yvec}{\mbox{$\vec{Y}$}}
\newcommand{\Pvec}{\mbox{$\vec{P}$}}
\newcommand{\pvec}{\mbox{$\vec{p}$}}
\newcommand{\Qvec}{\mbox{$\vec{Q}$}}

\newcommand{\vn}{\mbox{$v(\nvec)$}}
\newcommand{\vp}{\mbox{$v(\pvec)$}}
\newcommand{\gn}{\mbox{$\gamma(\nvec)$}}
\newcommand{\gA}{\mbox{$\gamma(\Avec)$}}
\newcommand{\Fc}{\mbox{$G(\cvec)$}}
\newcommand{\Gc}{\mbox{$G(\cvec)$}}
\newcommand{\FN}{\mbox{$G(\Nvec)$}}
\newcommand{\GN}{\mbox{$G(\Nvec)$}}

\def\bibdir{.}
\usepackage{boxedminipage}
\usepackage{graphicx}
\usepackage{float}

\newcommand{\scaledgetfig}[3]{
\begin{center}
\begin{figure}
\begin{center}
\includegraphics[keepaspectratio=true,hiresbb=true,width=#1\textwidth]{#2}
\end{center}
\caption{\sf #3}
\label{fig:#2}
\end{figure}
\end{center}
}

\begin{document}  
\bibliographystyle{unsrt}
\small
\noindent
\emph{This paper was published as:}\\
METALLURGICAL AND MATERIALS TRANSACTIONS A-PHYSICAL METALLURGY AND MATERIALS SCIENCE 1996, Vol 27, Iss 6, pp 1431-1440\\
\emph{and also in}
``The Selected Works of John W. Cahn,''
Edited W.C. Carter and W.C. Johnson, published by Minerals, Metals, \& Materials Society (December 1998) 
ISBN-10:087339416X,
ISBN-13:978-0873394161
\normalsize

\noindent
No pdfs available from the publisher.

\begin{center}
Crystal Shapes and Phase Equilibria:\\
A Common Mathematical Basis\\
\vspace{.5in}
J. W. Cahn, and W. C. Carter\\
MSEL, NIST, Gaithersburg, MD 20899\\
\vspace{.25in}
{\em{Submitted to Met. Trans. A, March 1995\\
for the special issue honoring Hub Aaronson\\
"Atomistic Mechanisms of Nucleation and Growth in Solids"\\}}
\vspace{.5in}
\begin{abstract}
Geometrical constructions, such as the
tangent construction on the molar free energy for determining
whether a particular composition of a solution, is stable, are related to
similar
tangent constructions on the orientation-dependent interfacial energy for
determining stable interface
orientations and on the orientation dependence of the crystal growth rate
which
tests whether a particular orientation appears
on a growing crystal.  
Subtle differences in the geometric constructions for the three fields
arise from the choice of a metric (unit of measure). Using results from
studies of extensive
and convex functions we demonstrate that there is a common mathematical
structure for these three disparate topics, and use this to find new uses for
well-known graphical methods for all three topics.
Thus the use of chemical potentials for solution thermodynamics is very
similar to
known vector formulations for surface thermodynamics, and the method of
characteristics which tracks the interfaces
of growing crystals; the Gibbs-Duhem equation
is analogous to the Cahn-Hoffman equation.
The Wulff construction for equilibrium crystal shapes can be
modified to construct a ``phase shape'' from solution free energies that is a
potentially useful method of numerical calculations of phase diagrams from
known
thermodynamical data.
\end{abstract}
\end{center}

\section{Introduction}

Hubert Aaronson's wide ranging contributions to
materials
science over
the last
four decades has focussed repeatedly on three major topics and their
applications to phase
transformations:\cite{HAust,HPT,HSS1,H53,H105,H145,H251}

1) solution thermodynamics and multicomponent phase
equilibria;\cite{H27,H28,H26,H133}

2) equilibrium shapes of surfaces with anisotropic surface
energy;\cite{H90,H93,H96,H98,H99,H103,H119,H128,H135,H279}
and,

3) the morphology of growing precipitates.\cite{HMass,H18,H23,H72}

In this paper, we hope to honor Hub and his contributions by
illustrating
how each of these topics derives from a common mathematical basis and
that the graphical constructions derived for each separately can
shed new insight into the others.

Tangent constructions are powerful graphical methods, widely
used for heuristic, computational and theoretical purposes in our
field.\cite{Gibbs1,Baker,HPT}  We will look at differences in how tangent
constructions are
used in each of three topics, and explore how these varied methods may be
used in the others.

For multicomponent phase equilibria (at constant temperature and
pressure) the tangent construction is performed on a plot of the molar
Gibbs free energy, $\Fc$, which has to be convex from below at equilibrium. 
(Here $\cvec = (c_1, c_2,...)$ is a short-hand [vector] notation for
the
molar composition of a multicomponent phase or system.)   For first order
phase changes, $\Fc$ may be multivalued with a different $\Fc$ for each
phase, usually with a different symmetry.  Concavities in this
plot
represent metastable and unstable phases and compositions for which the free
energy is too high. Such concavities lead to ranges in composition
(sometimes called miscibility gaps) where single phases are not in
equilibrium.  These concavities
are removed with the common tangent
construction, that identifies for a particular average
composition whether a single phase is in equilibrium, and if not what mix of
phases will have the lowest free energy.
This
graphical process, termed {\em{convexification}}, leads to a convex hull of
$\Fc$.  Each point on that convex hull represents the lowest free
energy of the system of a given average composition, including permitting the
equilibrium to be
multiphase.  Common tangents are often used to construct phase
diagrams from a nonconvex $\Fc$.

A quite similar set of graphical procedures is applied to the
orientation dependent surface free energy per unit area $\gn$, except
that the construction is performed on a radial plot of the {\em{reciprocal}}
of
$\gn$   
\cite{FCF63} .  Here $\nvec = (n_1,n_2,n_3)$ is the normal to the
surface.
Note
that the components of $\nvec$ are the analogs of the concentrations
of
the chemical species.\cite{Physique}  Concavities are removed by
convexification, and
the points on the convex hull of $1/\gn$ represent (the reciprocal of) the
lowest free
energy a surface with a certain average orientation can achieve. 
Interface orientations for which $1/\gn$ lies inside the convex hull
have too high an energy and are unstable. The tangent construction
identifies which surfaces become corrugated at  equilibrium, and
specifies which orientations of lower energy coexist to replace a high
energy surface, even though there is greater surface area.\footnote{The
tangent
plane to the reciprocal of $\gn$ is equivalent to a more awkward tangent
sphere
construction to $\gn$ itself due to Herring\cite{Herring} that was the first
stability test for surfaces.}  If there is more than one possible
{\em{phase}} state of the surface (facetted, surface melted, wetted, etc.)
$\gn$ may be multivalued, just as $\Fc$.  The convexification of $1/\gn$ has
many analogies to that for $\Fc$: It identifies the occurrence of orientation
gaps and surface phase transitions.

These two examples are based on finding minima in free energy.  For
the kinetic example we consider the cases of growth rates
$\vn$ that are constant in time and may be orientation
dependent.\cite{H64}  This occurs not only with interface controlled
growth\cite{TCH} and massive transformations,\cite{H46} but also with
such diffusional growth processes as cellular
precipitation,\cite{H36} eutectic and eutectoid growth,\cite{HAust}
discontinuous coarsening,\cite{JDL,CTH} liquid film migration
\cite{Yoon} and
diffusion induced grain boundary migration.\cite{CAH}  When there is
growth anisotropy, certain orientations tend to disappear from the
shape.  To
determine whether
a particular orientation will be part of a limiting outward growing shape,
the
same
graphical convexification is performed on a plot
of $1/\vn$.\cite{CTH}  Those fast growing orientations that will
eventually
disappear
from a growing crystal show up as concavities in this plot.  The
common tangent
construction will determine whether some orientations will
disappear into an edge or a corner depending on whether the
tangent plane touches $1/\vn$ at two or more distinct points.
No energy minimization is involved, but it is a Huygens principle of
least time.\cite{CTH,Gelfand}

In these constructions there are many analogies.  Composition gaps
have their analogs in orientation gaps; two phase equilibria become
edges, three or more phases in equilibrium become corners.  There are
quite analogous conditions on the curvature of $\Fc$ and $1/\gn$ for
stability with respect to undulations in $\cvec$ and $\nvec$; both are
called spinodals.\cite{Mullins}

There is another well known graphical construction that confirms the
analogy between the orientation dependence of $\gn$ and its kinetic
counterpart $\vn$. The Wulff construction performed on $\gn$ gives the
shape with the least surface energy for the volume it
contains.\cite{Taylor78} 
The
same construction on $\vn$ gives the limiting shape of a growing
crystal; it also the shape that will grow most slowly, the one that
adds the
least volume.\cite{TCH}
The Wulff shape is more basic than the convexified $\gn$. It contains all the
information that is in a convexified
$1/\gn$, but the converse is not always true,\footnote{For low symmetry
crystals the Wulff shape is unique, even
though $\gn$ can not be uniquely determined.\cite{H103,Arbel75}} and it is
more
convenient than $\gn$ for many applications.\cite{wmc}. The following
question
suggests itself:
Is there an equivalent Wulff-like construction for $\Fc$?
As we will show below, the answer is {\em{yes}} and the construction gives a
new method for obtaining a well-known plot in solution thermodynamics.

In this paper we will briefly describe the mathematical basis that
links all
three topics. A thorough discussion of this topic will appear
elsewhere.\cite{AMM1} Because theoretical thoughts about these topics
developed
quite
independently, exploration of these analogies creates opportunities to
exploit
the various separate methods and discoveries for new uses. We will try
to
answer how far these analogies can be pushed, and which methods
developed for
one of these topics can be adapted to the others. 

One example has
already been suggested and put to use.\cite{Physique}
Information about stable compositions is efficiently
stored in the phase diagrams in which simple rules derived from solution
thermodynamics and the phase rule play an important role in their construction,
interpretation and in many applications.  These diagrams identify
stable compositions, two-phase regions with tie-lines, three-phase
tie-triangles, etc., joining coexisting compositions.  The phase rule allows
a
cataloging of first-order phase changes.
Phase diagram extrapolations are a very useful tool for predicting stability,
metastable equilibrium
compositions, and the order in which phases appear upon cooling.
Information about stable orientations can be stored in an analogous
diagram,
called
an $n$-diagram 
where interface orientation play the same role as composition
in a phase diagrams.  The orientations that meet at each point on a curved
edge
are joined by tie lines; tie polygons specify the orientations that meet at
corners.  First-order surface phase changes, such as wetting and facetting,
conform to a modified phase rule.\cite{Physique}

However, perhaps these analogies are not so direct as they seem as
some puzzles should have arisen in the minds of the reader.
Namely:
 \begin{enumerate}
          
      \item Why are the tangent constructions performed on
        $\Fc$ while they are performed on $1/\gn$ or $1/\vn$?

      \item The condition for local stability for a
               two component systems is   $\Fc'' > 0$, while the
               equivalent condition, $\gn + \gn'' > 0$, on
               two-dimensional crystals is more complex.  For more than two
               components the Hessian, the matrix of second derivatives of
               $\Fc$, must be positive definite for phase stability, while the
               condition on $\gn$ for three dimensional surfaces can not be so
               simply expressed.  Are there formulations in which equivalent
               conditions have the same simple form?  

      \item Energy is minimized for $\Fc$ and $\gn$;
                what is minimized for $\vn$?

   \end{enumerate}

These conundrums will disappear when these topics are put into a single
mathematical framework.

\section{Convexification, Common Tangents, and Phase Diagrams}

We begin by {\em{extending}} the functions $\Fc$, $\gn$, and $\vn$ from
quantities
which are refer to chemical energy per {\em{unit}} mole, surface
energy
per {\em{unit}} area, and distance traveled per {\em{unit}} time to
quantities
which refer to the `free energy' of a system containing a specified
number of moles or a surface with an specified area, or the `distance'
that an interface has moved in a specified time.
In thermodynamics these are called extensive variables.\cite{Denbigh}
In the mathematics literature such functions are called
{\em{homogeneous degree one}} (HD1) 
\cite{Gelfand}
and are defined by the property:
\begin{equation}
        \label{eq: HD1}
H(\lambda \xvec) = \lambda H(\xvec)
\end{equation}
\noindent We limit ourselves to positive homogeneity where $\lambda$ is a
real and positive scalar.

When we homogeneously extend the three functions ($\Fc$, $\gn$, $\vn$),
by
letting $1/\lambda$ be the number of moles $\|\Nvec\|$, the area $\|\Avec\|$,
or the time $\|\pvec\|$, we
obtain:
\begin{eqnarray}
      \label{eq: extensions}
\FN  = \|\Nvec\| \Fc = \|\Nvec\| G(\Nvec/\|\Nvec\|)\\
\gA = \|\Avec\| \gn \nonumber = \|\Avec\| \gamma(\Avec/\|\Avec\|)\\
\vp = \|\pvec\| \vn = \|\pvec\| v(\pvec/\|\pvec\|) \nonumber
\end{eqnarray}
where $\|\Nvec\|$ is the number of moles, $c =\Nvec/\|\Nvec\|$, and $\Avec$
is a vector that represents a surface.  Its direction is
along the outward normal $\nvec$ and its length is the
area $\|\Avec\|$;  $\Avec = \|\Avec\| \nvec$, or $\nvec= \Avec/ \|\Avec\|$. 
Its components
$(A_1,A_2,A_3)$ are the projected areas along the three coordinate
axes.
$v(\pvec)$  represents the
distance the interface with orientation $\nvec$ will travel in time $\|\pvec\|$
in
a direction parallel to $\nvec = \pvec/\|\pvec\|$.

Note that $\FN$ is the familiar extensive function from
solution thermodynamics.  We do not change the symbol for the
functions, e.g.
$\FN$ and $\Fc$ are the $same$ function, but the later is
restricted to a restricted set ($\|\Nvec\| = 1$) of the space
of systems of all sizes and compositions, parameterized by $\Nvec$. 
Note also that $\|\Nvec\| = (N_1 + N_2+ \dots N_m)$, where $N_i$ is
the amount of component $i$, while $\|\Avec\| = \sqrt(A_1^2 + A_2^2 +
A_3^2)$.  The difference in form between these two expressions
will be seen to have important consequences.

The three extended functions $\FN$, $\gA$, and $\vp$ are actually what one
would infer from a particular experiment.
For instance: in a calorimetry experiment, the enthalpy of
a closed system is measured, but the value is reported as what
would have been measured if one mole (or one Kg) were present; the
surface
free energy is unlikely to be measured for a square meter, but is reported that
way;
the positions of a moving surface are rarely measured at one second intervals.

Any HD1 function is fully determined if its value is
known along some curve which intersects all rays emanating from
the origin.
We use this property in equation (\ref{eq: extensions})
to extend $\Fc$, $\gn$ and $\vn$ homogeneously to vectors of arbitrary
magnitude and to compute their values on the plane $\|\Nvec\| = 1$, and on
the spheres $\|\Avec\| = 1$, and $\|\pvec\| = 1$.
Another way an HD1 function can be reconstructed is from one of their
level sets, i. e. the set of points $\xvec(c_1)$ for which $H(\lambda \xvec)
= \lambda H(\xvec) = c_1.$  Then $H(\xvec) = c_1
\|\xvec\|/\|\xvec(c_1)\|$ where $\xvec$ and $\xvec(c_1)$ are in the same
direction.

The gradients of any HD1 function $f(\Xvec)$ depend only on the
direction of $\Xvec$, but not its magnitude, $\nabla H(\Xvec) = \nabla
H(\lambda \Xvec)$.  For $\FN$ this gradient is a vector; since $\partial
F/\partial N_i = \mu_i,$ 
the $i^{th}$ component of this vector is the chemical
potential of the $i^{th}$ species.  Consistent with this principle,
chemical potentials depend only on the composition.  

Any HD1 function can be written as the dot-product of its gradient
and its argument vector, and its argument vector is perpendicular to
the differential of its gradient:
\begin{eqnarray}
    \label{eq: dot_rep}
H(\Xvec) =  \Xvec \cdot \nabla H(\Xvec)  \\
0  =  \Xvec \cdot d\nabla H(\Xvec) \nonumber
\nonumber
\end{eqnarray}
For $\FN$ these are the familiar integral expression for the Gibbs
free energy $\FN = \Sigma_i (N_i \mu_i)$ and the Gibbs-Duhem
equation\cite{Denbigh}
$ \Sigma_i (N_i d\mu_i) = 0 $.  The gradient
of $\gA$, called  the vector $\xivec$,  has these properties, which
have been used for anisotropic surfaces.\cite{HoffC,CHoff}  The
gradient of $\vp$ is
the characteristic of the motion of the surface.\cite{FCF58,FCF72,CTH}

We next review the mathematics of convex functions.\cite{Gelfand}  A scalar
function $f$ of $\nu$ variables, or of a $\nu$-dimensional vector, is said to
be convex if it is bounded
from below, it is not everywhere infinite, and if
\begin{equation}
  \label{eq: convexity}
f(\lambda \Pvec + (1 - \lambda) \Qvec) \leq
   \lambda f(\Pvec) + (1 - \lambda) f(\Qvec)
         \; \; \; \; {\rm{for}} \; \; 0 \leq \lambda \leq 1
\end{equation}
If $f$ is also HD1 this inequality can be simplified.  Setting $\Xvec =
\lambda
\Pvec$ and $\Yvec = (1 - \lambda) \Qvec$, and
making use of Eq. \ref{eq: HD1} gives:
\begin{equation}
  \label{eq: convexity_HD1}
f(\Xvec + \Yvec) \leq f(\Xvec) + f(\Yvec)
\end{equation}

The definitions in Eqs. \ref{eq: convexity} and \ref{eq: convexity_HD1}
can be extended to partitions of vectors into a sum of an arbitrary
number of terms: i.e., $f(\sum_i X_i) \leq \sum_i f(X_i)$ for
a convex HD1 $f$.

\subsection{The Basis for Convexification}

In this section we show that the functions defined in Eq. \ref{eq:
extensions}
must be convex in the kind of minimizations that are representative of
thermodynamic equilibrium.  It is apparent from equation
(\ref{eq: convexity_HD1}) that convexity applies to the $\FN$ of any
chemical system.  If any two chemical systems are combined, the
masses of their individual 
components are added; this is equivalent to a vector addition of the
$\Nvec$ as $\Xvec$ and $\Yvec$ are added in equation (\ref{eq:
convexity_HD1}).  But the equilibrium free energy of the combined
system can not be greater than the sum of the equilibrium free
energies of the separated systems.  If the two systems remain unmixed,
the resultant free energy would be the sum of the free energies of the
parts; \footnote{The reason convexification need not apply to elastically
coherent systems is apparent when one considers that a two phase coherent
system can have an free energy that is the sum of the free energies of the
separated phases {\em{plus}} the elastic energy to make them
coherent.\cite{simple}} any
relaxation towards equilibrium can only lead to a reduction in free
energy.  Thus with the use of equation (\ref{eq: convexity_HD1}), the
convexity of $\FN$ is a simple consequence of
thermodynamics.

We next show with reasoning that is quite similar that convexity also applies
to $\gA$.  The convexified $\gA$ is the lowest free energy that a surface
with
a
planar perimeter with orientation $\nvec$ and spanning an area $\|\Avec\|$
can
achieve, allowing facetting to all other orientations.
Note that adding two area vectors, $\Avec_a$, $\Avec_b$, gives a another
area vector, say $\Avec_c = \Avec_a + \Avec_b$, which lies in the plane
spanned by $\Avec_a$ and $\Avec_b$.
This allows a simple construction for the addition of area vectors.
Let the area vectors be represented by rectangles of area, $\|\Avec_a\|$,
$\|\Avec_b\|$, and $\|\Avec_c\| = \|\Avec_a + \Avec_b\|$.
Since the normals to these three rectangles lie in a plane, the three
rectangles form a `tent' (or, triangular prism) and we will take the
rectangle representing the summed area $\|\Avec_c\|$ as
the `tent floor.'
\footnote{
Note that there are additional areas associated with the two triangles
at the front and back of the tent, but these areas can be made negligible
by making the rectangle very long compared to its width or, equivalently
by corrugating the roof, keeping the orientations fixed, to form
a series of similar small tents, like a `factory roof'.}
The proof that $\gA$ is convex parallels the proof that $\FN$ is convex. 
Consider the area $\|\Avec_c\|$.  Its energy cannot exceed the energy of the
combined areas $\Avec_a$ and $\|\Avec_b\|$; if this were not true $\|\Avec_c\|$
would spontaneously form a tent. This must hold for all possible
configuration of `tent sides.'  But clearly $\gamma(\Avec_c)$ can be less than
the tent energy.  Thus $\gA$ is convex at equilibrium since
$\gamma(\Avec_c) \leq  \gamma(\Avec_a) + \gamma(\Avec_b)$ for all
$\Avec_a + \Avec_b = \Avec_c$.  Note that the magnitude of
area itself is a convex function; the combined area
cannot exceed the sum of the separate areas.

The consequences are also similar.  If $\gA$ is a convex function, then all
orientations are stable with respect to facetting.  Since $\gamma(\Avec_c) \leq
\gamma(\Avec_a) + \gamma(\Avec_b)$, 
formation of a tent or corrugation of a surface represented by $\Avec_c$ into
{\em{any}} a configuration represented by two other vectors that
sum to $\Avec_c$ cannot decrease the energy of the original structure.
Conversely, if $\gA$ is {\em{not}} convex at $\Avec_c$, then there must
be a corrugated structure which is composed of alternating pieces $\Avec_a$
and $\Avec_b$ which has a lower surface energy.
The same construction can be applied to the formation of corners by considering
a partition into three or more orientations.

It is important to note that this convexity criterion comes from
thermodynamics on a very general and fundamental level.\footnote{In the above
argument we have ignored the energies contributed by edges and corners
separating pieces of planar surface, just as we have ignored the energies of
surfaces between coexisting phases in minimizing $\FN$ by convexification. 
But in the surface case, if such other energies exist they can not be ignored
in forming the limiting factory roof.}   The
inequality (\ref{eq: convexity_HD1}) applies to $\FN$ and
$\gA$, and not to the molar free energy $\Fc$ or the surface
free energy per unit area $\gn$.  
The inequality (\ref{eq: convexity_HD1}) should not and does not apply
to $\Fc$. Note that $\Fc$ when convexified for equilibrium curves up
instead of down.  For a binary solution, comparing $G(c_a + c_b)$ with
the sum of $G(c_a)$ and $G(c_b)$ does not make any sense, since mass
is not conserved. \footnote{The free energy of a system with one mole
should not be compared to the sum of that of two others, each with one
mole.}  Although the solute species is conserved, the mass of the
solvent is not.

The inequality (\ref{eq: convexity}) applies to
$\FN$ and $\gA$, and to any planar submanifold (a lower dimensional planar cut,
including any straight line
section) of the extended functions of $\nu$ or $3$ dimensional variables.  When
$\Pvec$ and $\Qvec$ are taken as end
points of
vectors, the end point of the vector $\lambda\Pvec + (1 -
\lambda)\Qvec$ is always on the connecting straight line segment.
Thus $\FN$ is convex from below on any straight line section; this
includes the hyperplane (for which
$\|\Nvec\| = 1$) of molar free energies $\Fc$.  Thus convexity applies to
$\Fc$.  The widely used graphical
convexification methods for $\Fc$ are thus validated. 

Applying the inequality (\ref{eq: convexity_HD1}) for $\gA$ is valid;
but applying (\ref{eq: convexity}) makes little sense for $\gn$.  When
$\Pvec$ and $\Qvec$ are taken on the unit sphere, that is as end
points of unit vectors,
the end point of the vector
$\lambda \Pvec + (1 - \lambda)\Qvec$
is always on the connecting chord; the inequality,
while correct, applies to a vector that is not a unit vector, one that
is in the interior of the unit sphere, and thus not to $\gn$.
\footnote{One can create a meaningful inequality for $\gn$ by lifting
this HD1 function from the chord to the surface of the unit sphere, but
a simpler method is developed here.} Thus
whether the scaled functions $\Fc$, $\gn$, and $\vn$ when restricted
to
some submanifold are also convex depends on
the somewhat arbitrary choice of how a unit of the argument is
measured; in other words, whether the choice of the operation which is
implied
by the operator $\|\|$ is linear.

Another use of the inequality (\ref{eq: convexity}) is to note that the 
surfaces defined by
level sets of any convex function have to be convex when the
function is a positive constant and concave when the function is
negative.  Because $\|\Avec\|$ is the usual length of the area vector
and
plots as distance from the origin $r$, we can convert
the equation ($\gA = \|\Avec\| \gn = const$) for
the level surfaces for $\gA$ into radial plots of the reciprocal of
$\gn$.
Setting the constant to $1$ the equation for the level
surface becomes $r = \|\Avec\| = 1/\gn$.  Since $\gn$ is positive,
the radial plot of the reciprocal of $\gn$ has to be convex at equilibrium.  

Note that $\|\Nvec\|$ is not the usual length of a
vector, and is not the radial distance to the level set of $\FN$.  As
a result the convexity of an inverse plot of $\Fc$ has as little
significance as the convexity of a radial plot of $\gn$.  We next look into
the
definitions of the metrics of these quantities, to understand the basis for
these differences and to look for alternate definitions.

\section{Metrics}

In the previous section we noted that convexity applies to functions,
such as $\gamma$ and $G$, defined for all vectors, such as
$\Avec$ and $\Nvec$, rather
than restricting these vectors to unit vector, such as $\nvec$ to give
$\gn$,
for which $(\sum A^{2}_{i})^{(1/2)} = 1,$ and for the molar free
energy, $\Fc$, for which  $\|\Nvec\| = \sum | N_i | = 1$. The two expressions
for the unit vectors
are fundamentally different; area and the number of moles are examples
of different metrics.

Metrics that measure the distance of a point from the origin are
simple examples of a convex functions.
The most familiar metric for vectors (including the area vector
$\Avec$) is
the Euclidean
metric, also known as $L^2$, $\| \xbar \| = (\sum x^{2}_{i})^{(1/2)}.$
Convexity for this metric is just the triangle inequality; the length of any
side of a
triangle is not more than the sum of the lengths of the other two
sides.   $\Avec$ has this metric. 

Other metrics are appropriate in other physical situations.
Consider the driving distance between two intersections in
a city, the appropriate distance is the $L^1$ metric:
$\| \xbar \| = \sum | x_i |,$ sometimes called the Manhattan metric, since
such
distances apply to travelers who can
only travel on a rectangular grid, like the streets of Manhattan.
A vector, \Nvec\ in $R^{\nu}$ with $\nu$ components which are the
amounts (here taken to be number of moles) of each of the constituents
is an indication of the size of a chemical system.  The Manhattan
"length" of this vector is the total number of moles.  This length is
the factor $\lambda$ used to convert $\FN$ to $\Fc$.  

A simple way of extending the concept of metrics is to give different
weights to the components in the sums that define a metric.  A
weighted $L^2$ metric for area can account for some anisotropy, but we
know of no useful application.  The weighted Manhattan metric, occurs
quite naturally if mass and weight percent, rather than number of
moles
and mole percent, become the variables.  The mass of a system
is $\|\Mvec\| = \sum | m_i N_i |$, where $m_i$ is the molecular
weight of
the $i^{th}$ species.  

One limit of the weighted metrics, that gives zero weight to all but one of
the components, is used for both surfaces and chemical systems.  
This limit provides a link between the mathematics of surfaces and chemical
systems, and
permits other convexification methods to be used.   
For chemical systems, this weighting is used for molal concentrations,
the number of moles of solutes for a fixed amount of solvent (usually
one mole or one Kg). \cite{Denbigh} Molal concentrations are defined as $c^m_i
= N_i/N_1$ (or $N_i/m_1N_1$ with $m_1N_1 = 1 kg$, e.g. for aqueous
solutions).  The size of the system (length of the vector) is then
defined as the amount of solvent alone, regardless of the amounts of
the other components. Note that $\cvec^m \in R_+^{\nu - 1}.$  The unit length
is one mole or 1 kg.   The
molal free energy is $G({\cvec}^m) = G(1, c^m_2, c^m_3, \ldots )$  

Molal concentrations are on a special planar cut of the space of all $\Nvec$
and thus the
convexification applied to $G(\cvec^m)$ gives the same common tangents
and equilibria that would be obtained from $\Fc$. Molal free energies
also provide the same spinodal stability limits from the same
curvature criteria. 

For vicinal surfaces, area is often defined as the area
projected along some symmetry axis, giving no weight to other
components of the area vector.\cite{Andreev,L_and_L}
If we define the components of a
new orientation vector as $\nvec^m_i = A_i/A_1$ without regard to the
sign or size of this ratio, we have extended
the concepts of vicinal surfaces to all orientations, and the analogy with
molal concentrations is kept.  Note that $\nvec^m \in R^2$ lives in the space
of $\Avec$ on a planar cut perpendicular to one of the axes in the same way
as
molal concentrations do in the space of $\Nvec$.  We will denote the
surface free
energy per unit projected area projected along the $x_1$ direction as
$\gamma(\nvec^m) = \gamma(1,A_2/A_1,A_3/A_1)$.   Convexity applies to
$\gamma(\nvec^m)$.

Using an $L^2$ metric $(\sum N^{2}_{i})^{(1/2)}$ to describe the size of a
chemical system makes little physical sense, but, as we shall see, it opens
up
some useful surface techniques for chemical thermodynamics.  By defining a
Euclimolar metric $\|\Nvec^{Eu}\| = (\sum N^{2}_{i})^{(1/2)}$, we can define a
Euclimolar free energy
$G^{Eu} = \FN/\|\Nvec^{Eu}\| = \Fc/(\sum c^2_i)^{1/2}$, which is $\FN$ evaluated
on the unit sphere $\|\Nvec^{Eu}\| = 1$.

\section{Shapes from Gradient Construction}

We examine graphical constructions which follow from the
geometric relationships in Eq. \ref{eq: dot_rep} and give
rise to shapes which also demonstrate the correspondence between
our three topics.  These methods require that the extended function be
continuous and piece-wise differentiable ($C^1$), and therefore do not have
as
general an applicability as the methods associated with the Wulff
constructions.

\subsection{$\muvec$-Shapes from Solution Thermodynamics}

We illustrate the following example from two-component regular solutions
but the concepts certainly apply to more components and more sophisticated
solution models.  Although two components allows us to define a single
composition $c$, the comparisons are abetted by our introduction of a vector
notation.

For a regular solution model $\Fc = c_1 c_2 + T ( c_1 ln c_1 + c_2 ln c_2)$,
where $T$ is a reduced temperature which scales out the energy of mixing and
Boltzmann's constant.  Letting $c = c_1 = 1 - c_2$, the molar
free energy becomes $G_m(c) = c(1-c) + T (c ln c + (1-c) ln (1-c))$
In Figure \ref{fig:Fc_shape}, these molar free energies are plotted
in the left column of the figure at reduced temperatures above, just below
and
well below
the critical temperature.
The common tangent construction at each temperature was used to draw
the
phase diagram in the middle.
The dashed curve in the phase diagram correspond to the spinodals, where
$G_m''(c) = 0$.

\scaledgetfig{0.75}{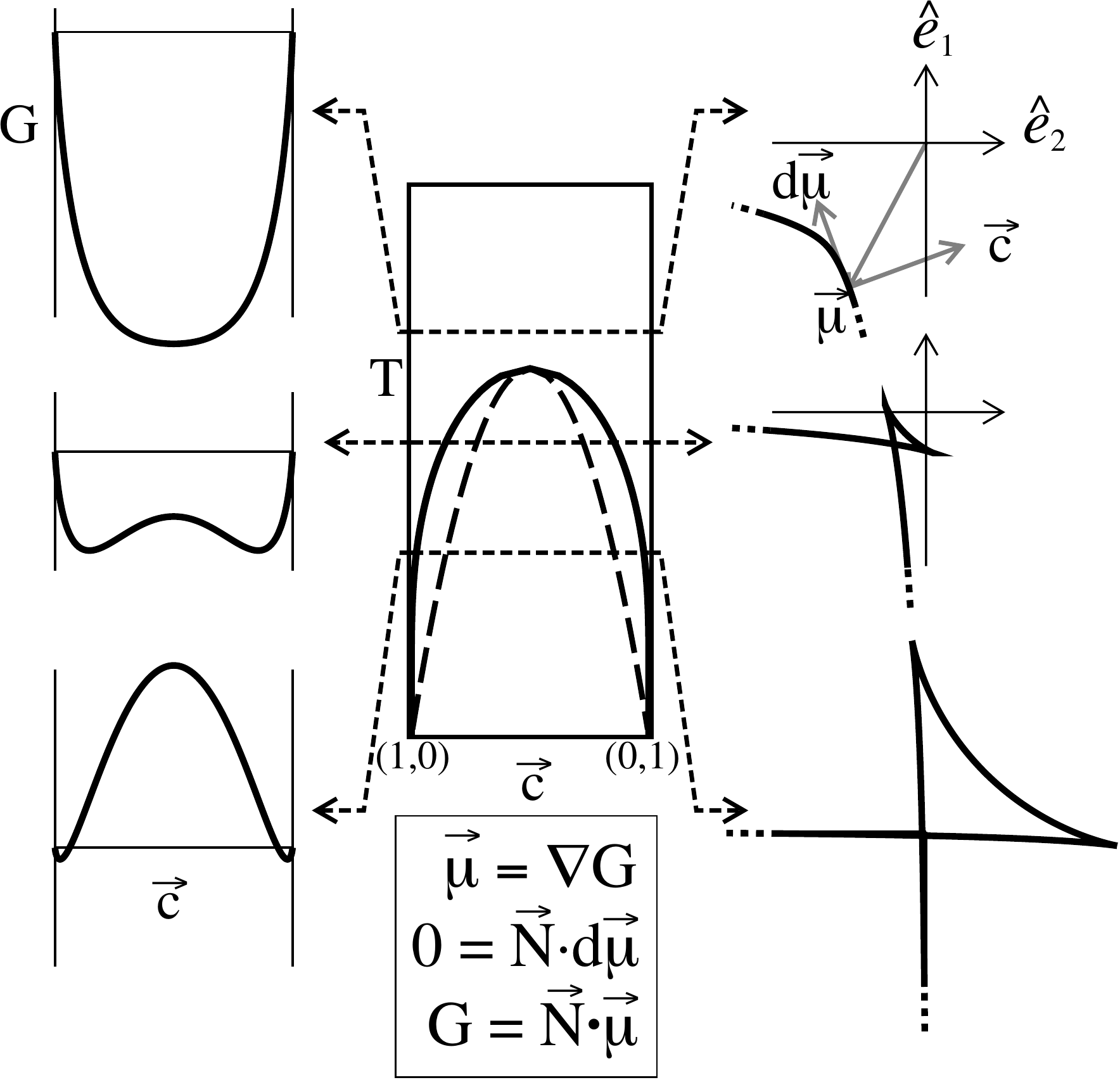}{In the left column molar free energies for
the regular solution
model at temperatures $1.1 T_{crit}$, $0.8 T_{crit}$, $0.5 T_{crit}$ are
plotted.  The chemical potentials at corresponding temperatures are
plotted on the right; these curves trace out the "$\muvec$-shapes."
The phase diagram is plotted for reference in the center.  
The crossings in the $\muvec$-shapes represent of two-phase equilibria. The
compositions $\cvec$ are given by the normals to curve $\muvec$ and
the two compositions in equilibrium at the crossing are also the
common tangent points which could be drawn on $\Fc$.  The ears represent
metastable and unstable compositions; the sharp points on the ears are the
spinodal points which are
represented by dashed lines on the phase diagram, or points
of inflection on $\Fc$.
}

Chemical potentials for both components can be obtained for this model by any
one of a number of equivalent ways, e.g. by taking the derivatives of $\FN =
(N_1 +
N_2)\Fc$ with respect to $N_1$ and $N_2$. Another is taking the intercepts at
$c = 0$ and $c = 1$ of tangents at $c$ to $G_m(c)$.  The resulting chemical
potentials are plotted against each other in the figures of the right column
of
Figure \ref{fig:Fc_shape}.  Since the coordinates of any point of this curve
give the values of the two chemical potentials, these point are the ends of
$\muvec$ is the gradient of $\FN$. We wish to focus on this $\muvec$-shape.

At high $T$ this shape is smoothly curved.   Because of the geometric
relation
in Equation \ref{eq: dot_rep} i.e., the Gibbs-Duhem equation in the case of
solution
thermodynamics, $\cvec \cdot d\muvec = 0$, the
normal to the $\muvec$ curve is the composition vector $\cvec$.  Thus we can
know the composition for each part of the curve.  Once that is known we can
recover $\Gc$ from this curve from $\Gc = \cvec \cdot \muvec$, but there are
other ways.

Below the critical temperature the $\muvec$-plot becomes self-intersecting
and
develops `swallow-tails' or `ears.'  The crossings are places
where two phases (smooth curves) have the
same chemical potential. Because of the Gibbs-Duhem equation relating slope to
composition, the distinct compositions of each of the phases
are given by the normals to the curve at the crossing point. The
sharpness the corner at the crossing relates to the difference in composition
between the two phases in equilibrium, i.e.,
the width of the miscibility gap in the phase diagram.
This analogy between corners in
$\muvec$-shapes and phase diagrams extends to
multicomponent phase equilibrium.  

The locally convex portions of the ears represent metastable compositions;
the
concave parts unstable compositions.  The metastable and unstable part are
separated by a spinode.  Eliminating the ears produces a convex figure that
is
the convexified $\muvec$-shape.  It contains all the information that was in
the convexified $\Fc$ plus a graphic display of all the phase equilibria.
This diagram illustrates the geometric nature of Equation \ref{eq:
dot_rep}.


The diagrams on the left and right side of Figure \ref{fig:Fc_shape}
are dual
to each other.
Each can be used to calculate the phase diagram in the center and any
diagram
on the right side ($\Fc$) can be used to calculate its dual
$\muvec$-shape
which appears on the right side of Figure \ref{fig:Fc_shape}.
Note that even though the phase diagram cannot be used to determine 
any of the other figures uniquely, the special CALPHAD procedures have
had
considerable successes.\cite{Kaufman,H145}

\subsection{$\xivec$-plots}

To draw out the analogy to the above discussion of binary phase
diagrams,
we consider an example of a two dimensional crystal.
Discussion of three dimensional crystals can be found elsewhere\cite{AMM2}.

A parallel geometric construction is made for an orientation dependent
surface tension ($\gn = 1 + \alpha n_1^2 n_2^2$) in Figure \ref{fig:gn_shape}.
This particular example is a first order expansion of a $\gn$ with
square symmetry. It could also be written as
$\gamma(\theta) = 1 + \alpha (Cos^2 (\theta)Sin^2(\theta)$.
The figure is remarkably similar to the construction for $\FN$ in
Figure \ref{fig:Fc_shape} except that the figures are closed curves since
the
range of $\Avec$ is all of $R^2$

\scaledgetfig{0.75}{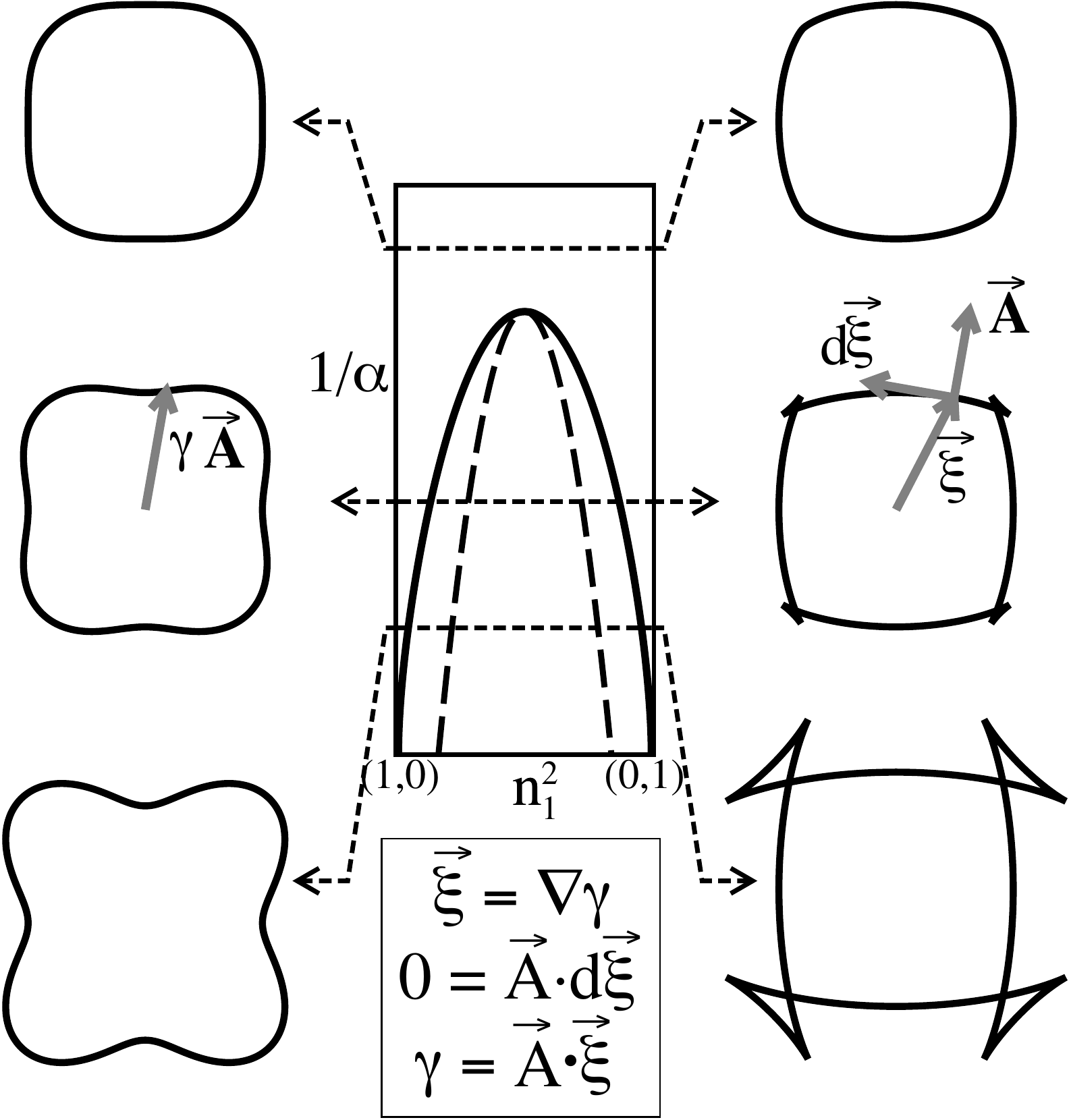}{
An analogous construction to that
in Fig. \ref{fig:Fc_shape} for an anisotropic $\gn$ for various values of an
anisotropy parameter $\alpha$ (see text).
In the left column $\gn$ is plotted from top to bottom for
$\alpha = 1/2$, 1, 2.
Anisotropy increases with positive $\alpha$, so $1/\alpha$ is
used in the $n$-diagram to correspond to the temperature
axis in Fig. \ref{fig:Fc_shape}.
The shape resulting from the gradient construction--with the
ears removed--is the surface of the Wulff shape.
}

Increasing values of $\alpha$ increase the anisotropy in
$\gn$ and
tends to create higher energy orientations which disappear from the
equilibrium
shape.
Thus, $1/\alpha$ plays a similar role to temperature, T, on the
construction for the regular solution $\Fc$ and so $1/\alpha$ is the
ordinate for the $n$-diagram illustrated in the center of
Figure \ref{fig:gn_shape}.
The critical value of alpha is 4/7.
Also, note that we use 
$n_1^2 = Cos^2(\theta)$
as the ordinate which is
convenient
for this case of square symmetry.

Plots of $\gn$ appear on the left side of Figure \ref{fig:gn_shape}
for
three different values of $1/\alpha$, one above and two below the
critical
anisotropy.
The gradient construction shown on the right column of the figure show that
`ears' develop as the anisotropy
increases
just as in the case for lower temperatures in the gradient
construction for $\FN$.
Any orientation on the `ears' is unstable and will break up into
orientations
given by the crossing in the $\xivec$-plot.
Those parts on the concave part of the `ears' (outside the
spinodes) are metastable.\cite{Mullins}

Consider the geometrical relations for the gradient construction in
Figure \ref{fig:gn_shape}.
According to Equation \ref{fig:gn_shape} (for surface energies, these
are the Cahn-Hoffman equations \cite{HoffC,CHoff} $0 = \Avec \cdot
d\xivec$) the
unit normal to the surface $\xivec$ must be the orientation vector.
Therefore, for all stable orientations, the surface of $\xivec$ must
also be
the surface of the Wulff shape.
In this sense, the interior region of the $\xivec$-plot must be
equivalent
to that obtained by the Wulff construction (See below).


\subsection{Growth Shapes and Method of Characteristics}

The method of characteristics has been used to integrate
the first order partial differential equations that are obtained for the motion
of a surface (or a growth front) when the velocity is a known function of
the surface orientation $v(\nvec)$. 
A thorough description may be found in Taylor and coworkers\cite{TCH,CTH} and
applications may be found in Carter and Handwerker\cite{CH93}.

Let $\tau(\xvec)$ be the arrival time of the surface at the position $\xvec$.
The level
set $\tau(\xvec) = t_{const}$ is the equation for the
position (or shape) of the surface at time $t = t_{const}$.
The gradient of $\tau$ is along the normal
of the level set and its magnitude must be inversely
proportional to the velocity: $\|\nabla \tau\| = 1/\vn$.
With $\pvec \equiv \nabla \tau$, the PDE is just a statement
that the HD1 function $v$ is a constant: $v(\pvec) = 1$.  The characteristics
are straight lines, given by the equation:
\begin{equation}
\label{eq: chars}
\xvec = \xvec_O + t \chivec(\nvec) \;,
\; \; \mbox{where} \; 
\chivec(\nvec) = \chivec(\pvec) = \nabla v(\pvec)
\end{equation}
and $\xvec_O$ is
the
surface at $t = 0$.

Letting the initial surface $\xvec_O$ be a point,
the calculation of the shape at a
fixed time (say t = 1) by the
method of characteristics gives the same result as the gradient formulations.

\section{Chemical Wulff Shapes}

There is a large literature regarding the Wulff shape that minimizes surface
energy for a given enclosed volume and the analogous kinetic Wulff shape that
give the limiting shape of a crystal growing outwardly under diffusion
control
that has recently been reviewed.\cite{TCH} The methods of construction are
the
same even though one is a minimization problem and the other is a long-time
solution of a first-order nonlinear partial differential equation.  

The method is one of iterative truncation of space by a set of half
planes--each half plane partitions the space into allowed and
disallowed half spaces.  For each value of $\nvec$
a plane is
drawn normal to $\nvec$ at a distance equal to the value of $\gn$ or $\vn$
respectively and the half space of all the more distant points discarded. 
When
this is done for all $\nvec$, the remaining points form a convex body that is
the Wulff shape.  The expression for the set of points which survive this
construction
is given by $\{\xvec | \xvec \cdot \nvec \leq \gamma(\nvec) \; \forall \nvec
\in S^2\}$
\noindent for $\gn$; substituting $\vn$ for $\gn$ gives the kinetic Wulff
shape.

The surface of the Wulff shape and the convexified $\xi$-plot
or plot of the characteristics are the same shapes, even though they are
obtained by quite different mathematical or graphical operations.  In the
Wulff
constructions there is no restriction to
either continuous or differentiable functions.  Because no differentiation of
date is used, the methods Wulff constructions may be quite superior for noisy
data.  Even though $\Fc$ is expected to be smooth for solutions it is
worthwhile to propose a Wulff construction for solution and compound free
energy data.

In order to do this we need to convert $\Fc$ into the Euclimolar free energy
$G^{Eu}(c) = \FN/\|N^{Eu}\| = \Fc/(\sum c^2_i)^{1/2}$.  For the binary
regular solution
example $G^{Eu}(c) = (c(1-c) + T (c ln c + (1-c) ln (1-c)))/(1 - 2c -
2c^2)^{1/2}$. The left hand panel of figure
\ref{fig:G_Eu_const} shows $\Fc$ for $T=0.45$ which
is $0.9$ of the critical temperature, and therefore shows a miscibility gap.
The second and third panels show $G^{Eu}(c)$, plotted respectively against
$c$
and as a radial plot, for this same temperature.  The third panel shows in
addition one step in the Wulff construction for a single composition. A line
for that composition is drawn perpendicular to another
line from the origin with slope $tan^{-1}(c/1-c)$ and length $G^{Eu}(c)$; all
points to the upper right of this line (shown gray) are discarded.  Performing
the Wulff
construction on a finite set of $c$ results in a fan of truncation lines
shown
in the last panel.
The
clear area at the lower left is the chemical Wulff shape, whose envelope is
the
$\muvec$-plot.  As with the $\muvec$-plot such a figure plots chemical
potentials against each other, and compositions are obtained from slopes.  It
identifies
single phases as continuous curves a two-phase equilibrium as the
corner. 

\scaledgetfig{0.33}{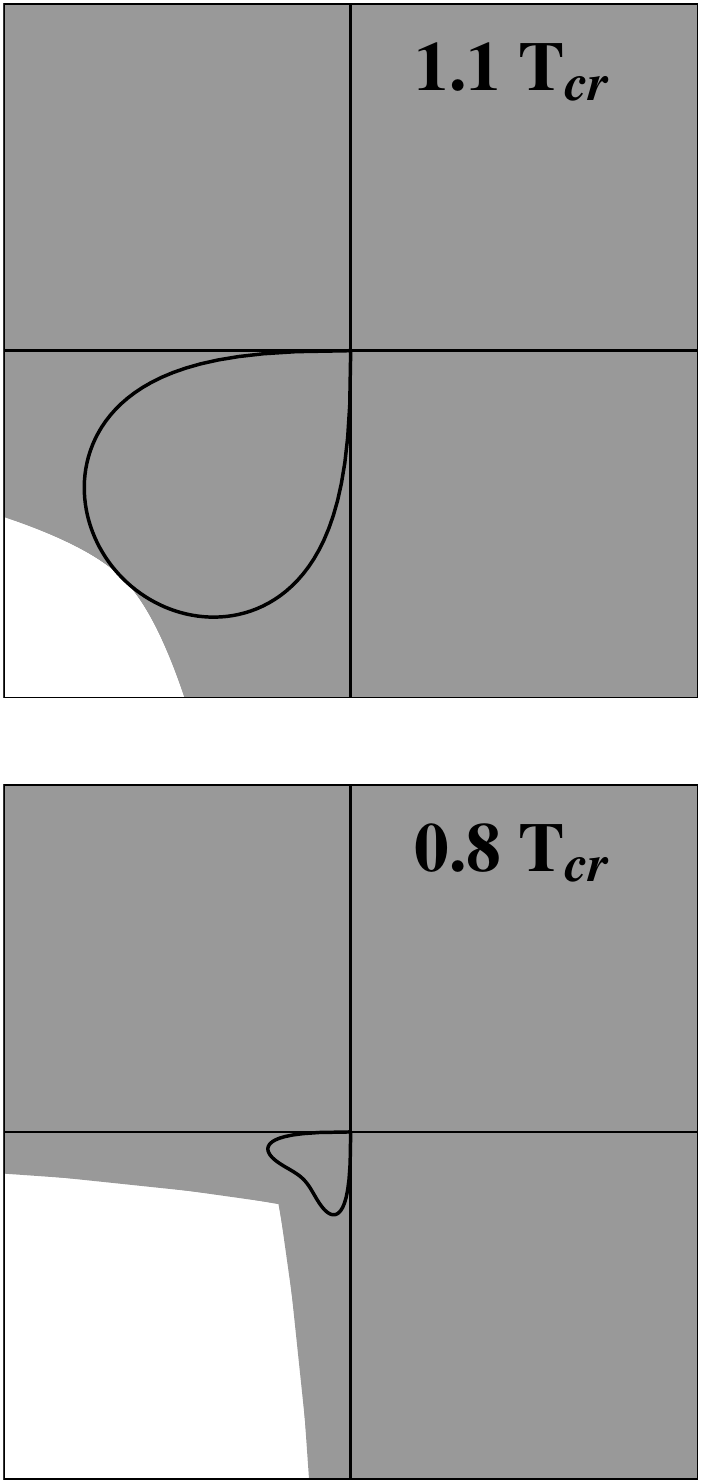}
{Illustration
of the chemical Wulff construction.
In the left figure, the molar free energy is plotted for a regular solution at 
$T = 0.9 T_{crit}$. In the middle two figures the Euclimolar free energy
$G^{Eu}(c)$ (see text)  for the same temperature
is graphed as heavy curves in standard format and radially as $\nvec
G^{Eu}(c)$. Note that the second and third plots look convex.
In the third panel, one step in the chemical Wulff construction is illustrated.
At a particular composition on $G^{Eu}$ a half plane is constructed
which is normal to radius (thin black line) drawn from the origin.
This divides the composition space into two parts: the gray region is to be
discarded.
In the final panel, the iterative elimination of discarded space yields the
chemical Wulff shape.
}

The inverse Wulff construction, finding the distance of a tangent line
from the origin
corresponding to a particular composition recovers the Euclimolar free
energy. Note that this is equivalent to $\GN = \muvec \cdot \Nvec$.

The concavity in the first panel of
Fig. \ref{fig:G_Eu_const}
shows that $\FN$ at this $T$ is not convex.  The
corner in the Wulff construction in the last panel confirms this.  Both are
appropriate criteria for nonconvexity of $\FN$.  The convexity of the curves
in the middle two panels is of
Fig. \ref{fig:G_Eu_const} are of no importance;
even though $\FN$ is not convex,
both curves are convex. 
For $\gn$ Herring's tangent sphere construction for finding stable
orientation
is an alternate test for convexity.  But this construction works only for
positive functions.  As can be seen in the third panel, it does not work for
a radial plot of a negative $G^{Eu}(\cvec)$.  

Note that the two-phase corner does not touch the Euclimolar free energy
plot;
the gap is a measure of the reduction in free energy upon phase separation.
The
chemical Wulff shape does not give metastable phases or their equilibria,
except when the entire curve of a stable phase is ignored in the
construction.
The undiscarded points in the interior of the lower left-hand area are not
physically realizable, except possibly as an unknown stabler
phase--an ice-9.\cite{CatsCradle} 

Examples of the chemical Wulff construction for the three temperatures
in Fig. \ref{fig:Fc_shape} are illustrated in
Fig. \ref{fig:G_Eu_exmpl}.  They show not only the miscibility gaps at the lower
temperature, but also the chemical potentials of the phases at various
compositions derived from the normals.
Note that in Fig. \ref{fig:G_Eu_exmpl} that the
Euclimolar free energy takes on some positive values as the
temperature is decreased.

\scaledgetfig{0.99}{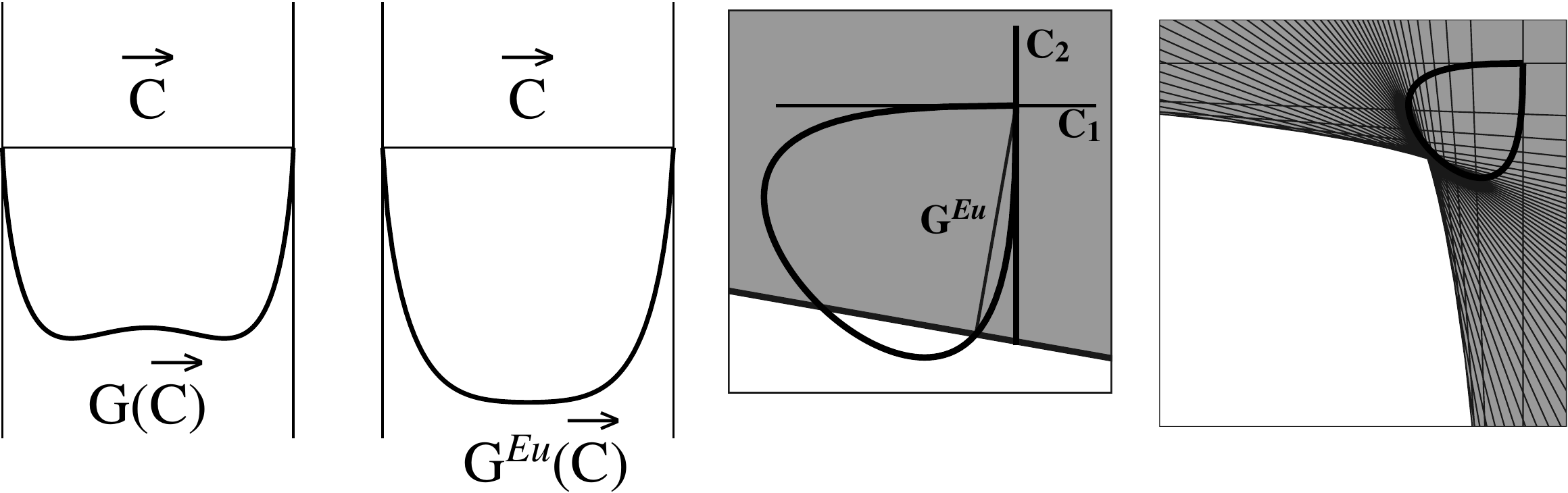}
{
Illustration
of the chemical Wulff construction for the three temperatures
which were used in \ref{fig:Fc_shape}.
The borders of the Wulff shapes are the same
as the $\muvec$-shapes without ears.
}

\section{Discussion}


\small
\begin{tabular}{|| c || c | c | c ||}
\multicolumn{4}{c}{Analogies Between Phase Equilibria and Shape
Morphology}\\
\hline \hline
    &  Composition Equilibrium & Surface Energy & Growth Shape\\
\hline
\hline
Convex&
    $\FN$ at Const. $P$ and $T$ &
         $\gA$ at Const. $\muvec$ and $T$ &
           $\vp$\\
Function & & &\\
\hline
Common Tangent&
      $\Fc$ &
        $1/\gn$ &
         $1/\vn$\\
Construction & & &\\
\hline
Gradient&
      $\muvec(\cvec) \equiv \nabla \FN$ &
        $\xivec(\nvec) \equiv \nabla \gA$&
           $\chivec(\nvec) \equiv \nabla \vp$\\
Formulation & & &\\
\hline
Geometric&
      $\Nvec \cdot d \muvec = 0$&
        $\Avec \cdot d \xivec = 0$&
          $\pvec \cdot d \chivec = 0$\\
Relations & & &\\
\hline
Wulff&
 $\{\muvec|\cvec \cdot \muvec \leq G^{Eu}(\cvec) \; \forall \cvec \in
\Sigma_{+}^{\nu-1} \}$ \footnotemark &
   $\{\xvec | \xvec \cdot \nvec \leq \gamma(\nvec) \; \forall \nvec \in
S^2\}$&
    $\{\pvec | \pvec \cdot \nvec \leq v(\nvec) \; \forall \nvec \in S^2\}$\\
Construction & & &\\
\hline
\hline
\end{tabular}
\normalsize

\footnotetext{$\Sigma_{+}^{\nu-1}$ is the simplex:
              $c_i \geq 0$ where
              $c_1 + c_2 + \ldots + c_{\nu} = 1$, in two-dimensions
           it is a line-segment; in three dimensions an equilateral
              triangle, etc.
              The chemical Wulff construction
              could as well have been written as
     $\{\muvec|\Nvec \cdot \muvec \leq \FN \; \forall \Nvec \in S_{+}^{\nu -1}
\}$,
           where $S_{+}^{\nu -1}$ the portion of the unit sphere embedded
              in $R_{+}^{\nu}$
              }

In this paper we have examined how the three topics to which Hubert
Aaronson has contributes so much, phase equilibria, shape equilibria,
and limiting shapes obtained with interface controlled growth, share a
common mathematical basis, which is that in each case there are HD1
functions that are to be convexified.  Because these topics have had
separate developments, there are many methods that have been found
useful in some but not all of the topics.  These fall broadly into
three areas as summarized in Table 1:  

\begin{enumerate}

\item Those that operate on sub-manifolds of the HD1 function, such as
$\Fc$, $\gn$, and $\vn$, and use such methods as finding its common
tangents and curvatures to give phase and shape diagrams, as well as
limits of metastability.  Here the choice of metric plays a role in
deciding which plot is to be convexified; and two choices for $\gn$
are
contrasted below.  The shape or $n$-diagrams are simple analogs of phase
diagrams with the missing orientations at edges and corners
represented as two and multi-phase regions, and coexistent
orientations represented by tie lines, triangles, etc.

\item  Those that operate on the gradient of the HD1 function, the
$\muvec$
and $\xivec$ plots and the characteristics to obtain shapes that have
ready physical interpretation for $\gamma$ and $v$.  Because of the
Gibbs-Duhem relation the normals to any point on the $\mu$ plot gives
its composition.  The surfaces of the innermost parts of this plot
represent equilibrium single phases and their composition ranges. 
Intersections to give edges and corners represent phases that are in
equilibrium with one another. The missing orientations at edges and
corners in this plot represent the composition gaps in the phase
diagrams, and limiting orientations at these edges and corners.

For the $\muvec$ and $\xivec$ plots the locally convex surfaces of the
``ears'' beyond these intersection represent metastable phases or
surfaces; all other parts of the ears are separated from the
metastable parts by spinodes and represent unstable phases or
surfaces.  

There is no clear cut stability criterion for the characteristics; any
characteristics can be stable at some time during shape evolution
with arbitrary initial data. 
However the limiting shape of an outward growing crystal is the
innermost plot, i.e. the plot without the ears.

\item Those that use a sub-manifold of the HD1 function in a graphical
construction to obtain a shape that for $\gn$ and $\vn$ is the Wulff
shape.  For $\gn$ this is the shape with the lowest surface energy for
the volume it contains; for $\vn$ this is the shape that for a given
volume would add the least volume by further growth; it is also the
limiting shape for outward growth.  Such a construction can be done
for $\Fc$, but the construction has to be modified because of the
Manhattan metric for $\Nvec$; this is awkward.  If we convert $\Fc$ to
a Euclimolar free energy, defined above, the unmodified Wulff
construction works to give a $\muvec$-shape with equilibria alone.

\end{enumerate}

The common mathematical basis has indeed made it possible to examine
the analogies for all three topics and for all three basic methods.  
The differences in methods, such as common tangents on radial plots of
$\Fc$ versus $1/\gn$, were often the result of the differences in the
metrics in conventional use.  Manhattan metrics work with the function;
Euclidian metrics with the reciprocal.  If we use a weighted metric
for $\gamma$, such as energy/(unit area) projected along an axis,
$\gamma^m$, the tangent constructions are done on this rather
than its reciprocal.  This metric is already in use for vicinal
surfaces.  Molar and molal free energies are convexified directly with
equivalent results.  The standard Wulff construction works with the
Euclidian metric, and neither molar or molal free energies are easily
used.  But with the definition of a Euclimolar free energy we can directly
create a $\muvec$ graphically, without taking derivatives of $\Fc$ or
$\FN$.

The analogies create many approaches for solving problems in all three
topics.  The advantages of having such flexibility in approach need to
be explored.  \cite{AMM1}

Phase diagram data are easy to obtain experimentally and can be obtained
without knowing the free energy.  Such data can be extrapolated. The
topology of phase changes is guided by the phase rule; such phase
changes appear on phase diagrams in standard formats.  The same holds
true for $n$-diagrams; the orientation of smooth surfaces and the
orientation gaps at edges and corners which develop at local
equilibrium, i.e., without waiting for full shape equilibration or without
measuring $\gn$.  From such data the $n$-diagrams can be constructed and
extrapolated, identifying surface ``phase changes'' that conform to a
phase rule that is modified for crystal symmetry.  We have in an
example \cite{AMM2} exploited this interconversion between shapes and
phase equilibria to analyze a complex series of phase changes in a
ternary regular solution.

All the information that is in $\FN$ is not only in $\Fc$, but also in
the $\muvec$ plots; the free energies can be recovered from such a
plot. 
The same interconversion holds for the chemical Wulff plots and the
convexified free energies.  These plots all display the same
information, but in different formats. Furthermore some plots are more
sensitive to errors in the data because differentiation or finding the
point of tangency is involved.  Which plot is best will be partially
determined by the nature of the data; chemical potential data ought to
go directly  into constructing a $\muvec$-plot.  Free energy plots
show
free energy and composition, but coexistent compositions must be
determined by a tangent construction that is very sensitive to errors in
the data and becomes increasingly difficult with increasing number of
components.  The Wulff shape is obtained without differentiating the
free energy data; the $\muvec$ shape requires taking a gradient of
$\FN$. 
These two plots should be congruent for the stable equilibria.  They
display phase coexistence clearly as corners and edges, compositions
as normal directions, which implicitly requires differentiation, and
free energy of a particular composition as the distance (in the
appropriate metric) of the corresponding tangent plane from the origin.

The analogies are not perfect, as a few examples will show.  The
stability criteria are different for the kinetics.  Curved surfaces
can be part of the Wulff shape, and thus of an equilibrium shape. 
Only points in the $\muvec$-shape represent equilibria.  Curved
surfaces
in this shape are ranges in the $\muvec$ and in the compositions.  A
system that spans such a range is not in equilibrium and has real-
space gradients of the chemical potential.  While there is a clear
analogy with edges and corners, there appear to be no chemical analogy
to a triple junction of surfaces.

\section{Summary}

In this paper, we have tied three fields together with a common
mathematics based on the fact that the underlying extensive function
must be convex.
From such a common basis, gradient constructions are derived which
have useful geometrical interpretations and allow results from
one field to be applied through analogy to the others.

Apparent differences in the way common tangents are applied to
compositions and to interfaces are resolved by consideration of the
particular metric in use.

The analogies lead to the notion of the chemical Wulff shape which
is constructed on chemical free energy normalized by the same
euclidian metric which is used to normalize surface tension.
This construction suggests a promising means to determine phase
boundaries without resorting to numerical differentiation.

Finally, the common mathematical structure presents a unified way
of studying, teaching, and understanding three important topics
in materials science.

\section{Acknowledgements}
This work was supported by the NIST Materials Science and Engineering
Laboratory,
Ceramics Division, and the Center for Theoretical and Computational Materials
Science.
We appreciate useful discussions with A. Roosen, J. Blendell, J. Warren, R.
Braun, J. Taylor,
C. Handwerker and Hub Aaronson.
We are also grateful to contribute to this acknowledgement of Hub Aaronson's
lasting contributions to our field.

\bibliography{\bibdir/hub_refs}
\end{document}